# UK General Election 2017: a Twitter Analysis

Laura Cram, Clare Llewellyn, Robin Hill
Neuropolitics Research Lab
School of Political and Social Science
University of Edinburgh
{laura.cram, c.a.llewellyn, r.l.hill}@ed.ac.uk

Walid Magdy
Institute for Language, Cognition & Computation
School of Informatics
University of Edinburgh
wmagdy@inf.ed.ac.uk

## 1. Introduction

This work is produced by researchers at the Neuropolitics Research Lab, School of Social and Political Science and the School of Informatics at the University of Edinburgh. In this report we provide an analysis of the social media posts on the British general election 2017 over the month running up to the vote. We find that pro-Labour sentiment dominates the Twitter conversation around GE2017 and that there is also a disproportionate presence of the Scottish National Party (SNP), given the UK-wide nature of a Westminster election. Substantive issues have featured much less prominently and in a less sustained manner in the Twitter debate than pro and anti leader and political party posts. However, the issue of Brexit has provided a consistent backdrop to the GE2017 conversation and has rarely dropped out of the top three most popular hashtags in the last month. Brexit has been *the* issue of the GE2017 campaign, eclipsing even the NHS. We found the conversation in the GE2017 Twitter debate to be heavily influenced both by external events and by the top-down introduction of hashtags by broadcast media outlets, often associated with specific programmes and the mediatised political debates. Hashtags like these have a significant impact on the shape of the data collected from Twitter and might distort studies with short data-collection windows but are usually short-lived with little long term impact on the Twitter conversation. If the current polling is to be believed Jeremy Corbyn is unlikely to do as badly as was anticipated when the election was first called. Traditional media sources were slow to pick up on this change in public opinion whereas this trend could be seen early on in social media and throughout the month of May.

## 2. Data Collection

A set of 56 keywords related to the British general election in 2017 (GE2017) was used to collect tweets on the topic. The Twitter streaming API was used to retrieve tweets containing any of these keywords between the April 29, 2017 and June 4, 2017. The keywords consist of hashtags, accounts, and terms representing phrases on the elections (e.g. #GE2017, general elections), politicians involved in the elections (Theresa May, Corbyn, #jc4pm), and related topics (e.g. Brexit, NHS).

During the period of study, over 34 million posts were collected, where 9.6 millions are tweets and 25 million are retweets. Figure 1 illustrates the volume of tweets/retweets collected daily during the period of study.

Figure 2 shows the average daily retweet versus tweet rate of the election related Twitter traffic. There appears to be a steadily increasing tendency to resend and reuse existing information as the election draws closer rather than to generate novel content.

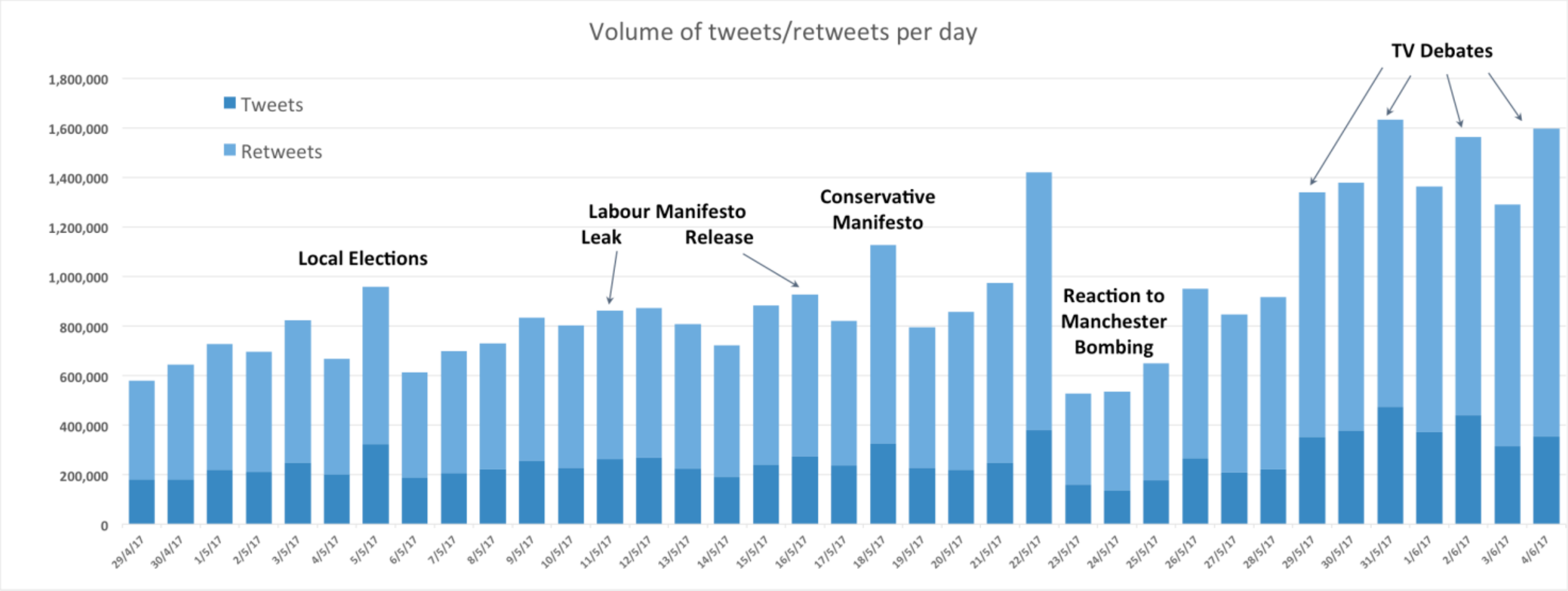

Figure 1. Distribution of the collected tweets/retweets over the period of study

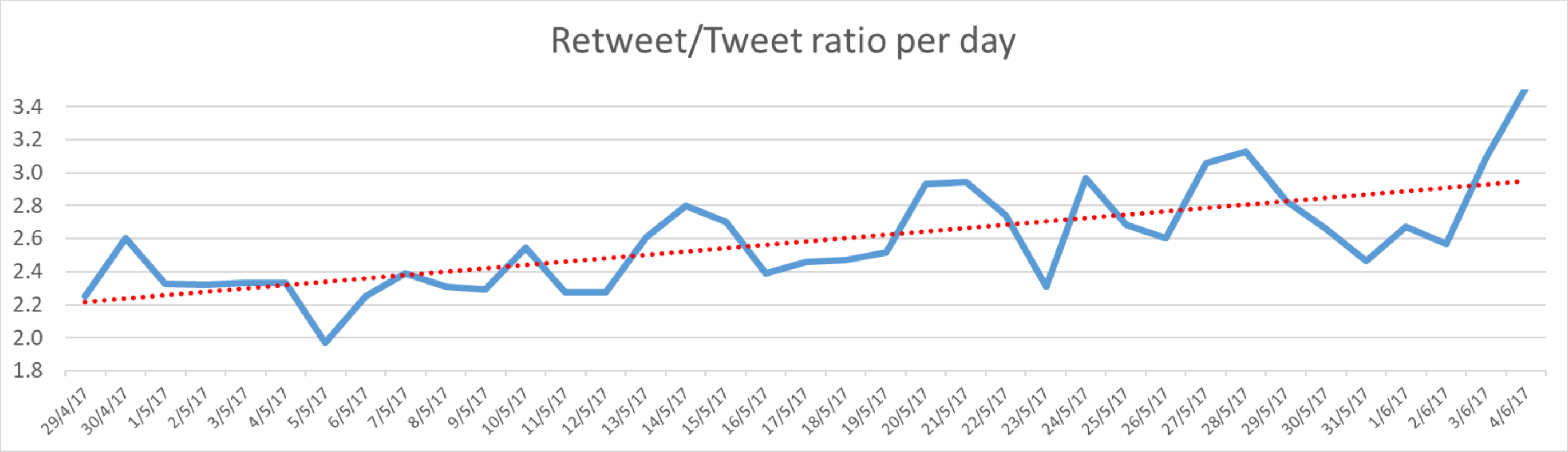

Figure 2. Relative rate of retweeting to tweeting each day. The red line indicates the linear regression or trend line (positive gradient).

## 3. Analysis

As shown in Figure 1, the number of posts collected per day, there is clear evidence of the event driven nature of the Twittersphere. As well as observing a rise in the overall number of posts related to GE2017 over the month, peaks in the data can readily be associated with events such as the 5 May local elections and the Manchester bombing (with an associated drop in GE2017 posting as the campaigns paused in its aftermath). Interestingly, the official release of the Labour manifesto did not produce a significant spike in overall Twitter posts, perhaps due its earlier leaking. However, as can be seen in Figure 4, it did enjoy two mini boosts (becoming one of the top three most used hashtags on the day of its leak and again on the day of its official launch). On 18 May the Conservative manifesto launch coincided with the ITV leaders' debate, producing a significant Twitter boost. Analysis of hashtags used, in Figure 4, indicates that the Conservative manifesto launch was indeed the larger contributor to this boost. As we are approaching the June 8 election date, Twitter traffic is evidently increasing. However, the increasingly event driven, and often top-down, shaping of the conversation is striking. Spikes in the Twitter data are closely linked with major media events and TV debates and, as we can see in Figure 4, are strongly influenced by the official hashtags promoted by the media companies.

## 3.1. Most popular hashtags over the collection period

The dominant role played by broadcast-driven and promoted hashtags is clear in Table 1 and Figure 3. Of the Top 20 categories of most employed hashtags, during the month preceding GE2017, the number two slot is occupied by those related to television and radio shows and the number five slot by the hashtags associated with the TV debates. In Figure 4, we see that these hashtags generated significant spikes, but prove ephemeral when contrasted with issues like Brexit which persisted throughout the campaign. There are very few substantive issues discussed in a sustained manner in relation to GE2017. Most of the Twitter traffic is pro or anti-leaders or political parties. However, of the issues discussed, Brexit dominates. It is the fourth most common hashtag employed (935,456) in the discussion of GE2017 and is employed more than twice as often as the next issue of significance, the NHS (420,092). The only other issues to feature in the top twenty hashtags over the full collection period are the potential further Scottish referendum (176,382) and the so-called dementia tax (154,007). The appearance of the Scottish independence question as the thirteenth most common hashtag employed in the discussion of GE2017 is particularly striking, given the UK wide nature of our data collection and the likely more localised interest in this issue in Scotland. This provides an indicator of the salience of this issue to those mobilised to tweet from Scotland.

Most striking in this data set is the overwhelming dominance of Labour tweeting. Tweets using hashtags do not necessarily indicate support but do highlight areas of discussion. With over one million tweets using Labour hashtags in our data set, Labour party coverage out-performs Conservative coverage by almost three times. There are no tweets employing anti-Labour hashtags in our top twenty most used hashtags collection. There are twice as many Labour (1,062,908) as Corbyn hashtags in the tweets (503,307). However, Corbyn hashtags, in position number six, still significantly outperform May hashtags (302,494) in position number nine. The pro-Labour momentum is boosted by the widespread use of Labour-promoted hashtags such as #forthemany, #forthemanynotthefew. Despite May's attempt to focus her campaign on her own leadership, rather than actively campaigning under the Conservative party banner, Conservative hashtags (381,647), at position number eight, marginally outrank May hashtags in the collection. Once again, however, the disproportionate presence of the Scottish National Party (SNP) in slot number eleven (244,481), given that only Scottish voters can elect this party, was striking. Interestingly it is the SNP, not the First Minister Nicola Sturgeon, that appears in the most common hashtag list. The Liberal Democrats, and their leader Tim Farron, do not figure in the top twenty hashtags. UKIP occupies slot nineteen (141,011), although Paul Nuttall does not appear.

Table 1: Top 20 Hashtag Categories Used throughout the Election Debate (the top 100 hashtags were grouped by topic and the top 20 topics selected)

| **Top Hashtags** | **Topic** | **Hashtags count** |
|---|---|---|
| #GE2017, #GE17, #GeneralElection, #GeneralElection2017, #Election2017 | General Election | 3,624,566 |
| #BBCQT, #marr, #Peston, #r4today, #NewsNight, #BBCSP, #VictoriaLIVE, #BBCDP, #WomansHour, #TheOneShow | TV/Radio | 1,195,880 |
| #VoteLabour, #Labour, #ImVotingLabour | Labour | 1,062,908 |
| #Brexit | Brexit | 935,456 |
| #BBCDebate, #BattleForNumber10, #ITVDebate, | Hustings / Debates | 844,514 |
| #LeadersDebate, #MayvCorbyn | | |
| #JC4PM, #Corbyn, #JeremyCorbyn | Corbyn | 503,307 |
| #NHS, #VoteNHS, #SaveOurNHS | NHS | 420,092 |
| #Tories, #Tory, #conservatives, #conservatives, #VoteConservative, #conservative | Conservatives | 381,647 |
| #TheresaMay, #May | May | 302,494 |
| #ForTheMany, #ForTheManyNotTheFew | For the Many | 271,251 |
| #voteSNP, #SNP | SNP | 244,481 |
| #ToryManifesto | Tory Manifesto | 214,973 |
| #ScotRef, #indyref2, #Scotland | Scottish referendum | 176,382 |
| #ToriesOut | Tories Out | 172,380 |
| #RegistertoVote, #Vote, #WhyVote, #Register2Vote | Register to vote | 165,757 |
| #Manchester, #Londonattacks, #LondonBridge, #London | Terrorist Attacks | 154,497 |
| #DementiaTax, #Socialcare | Social Care | 154,007 |
| #LabourManifesto | Labour Manifesto | 149,731 |
| #UKIP | UKIP | 141,011 |
| #BBCElection, #BBC | BBC | 124,755 |

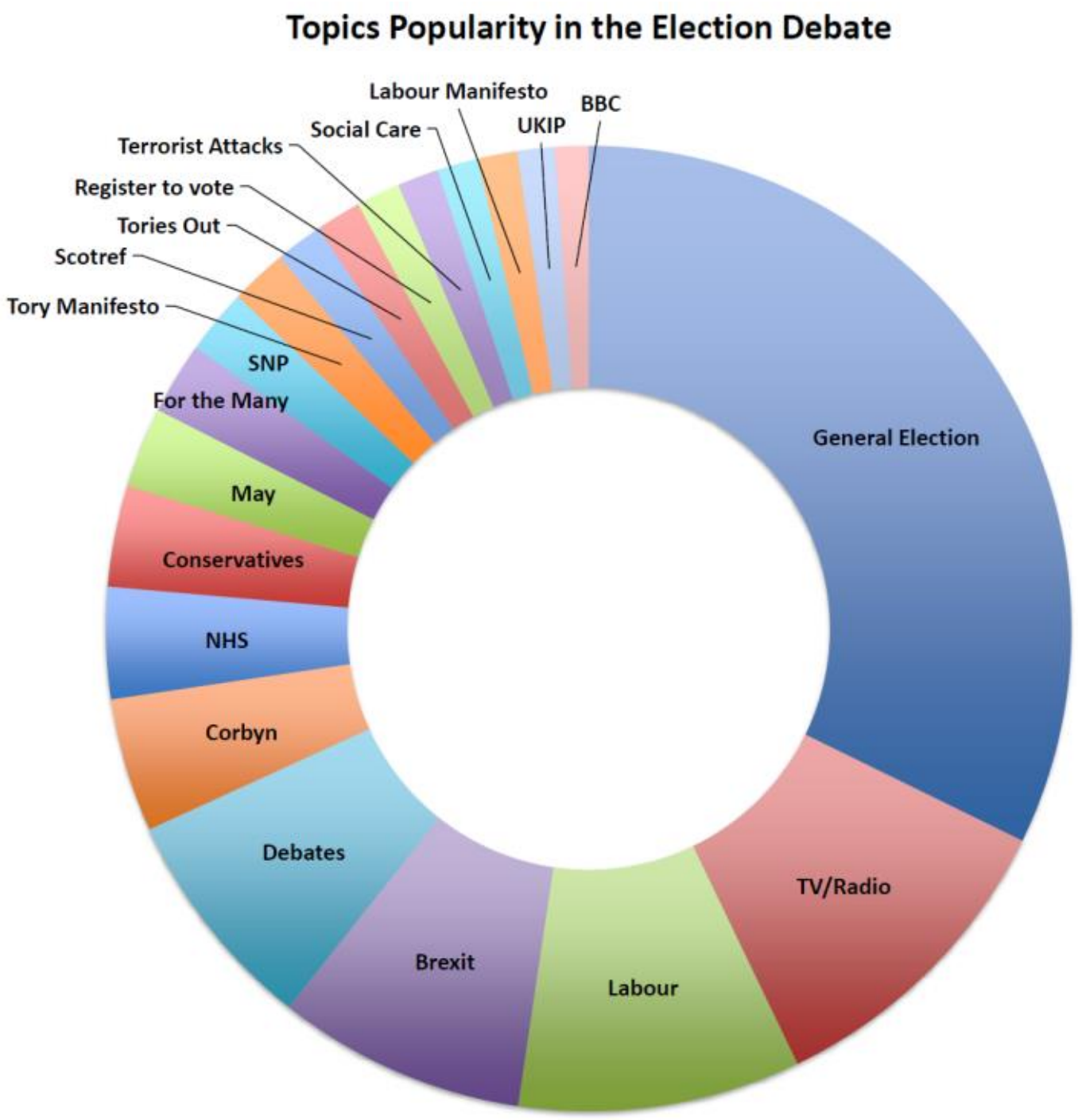

Figure 3: Percentage Share of Topics of the Top 20 Hashtags Used in the Election Debate

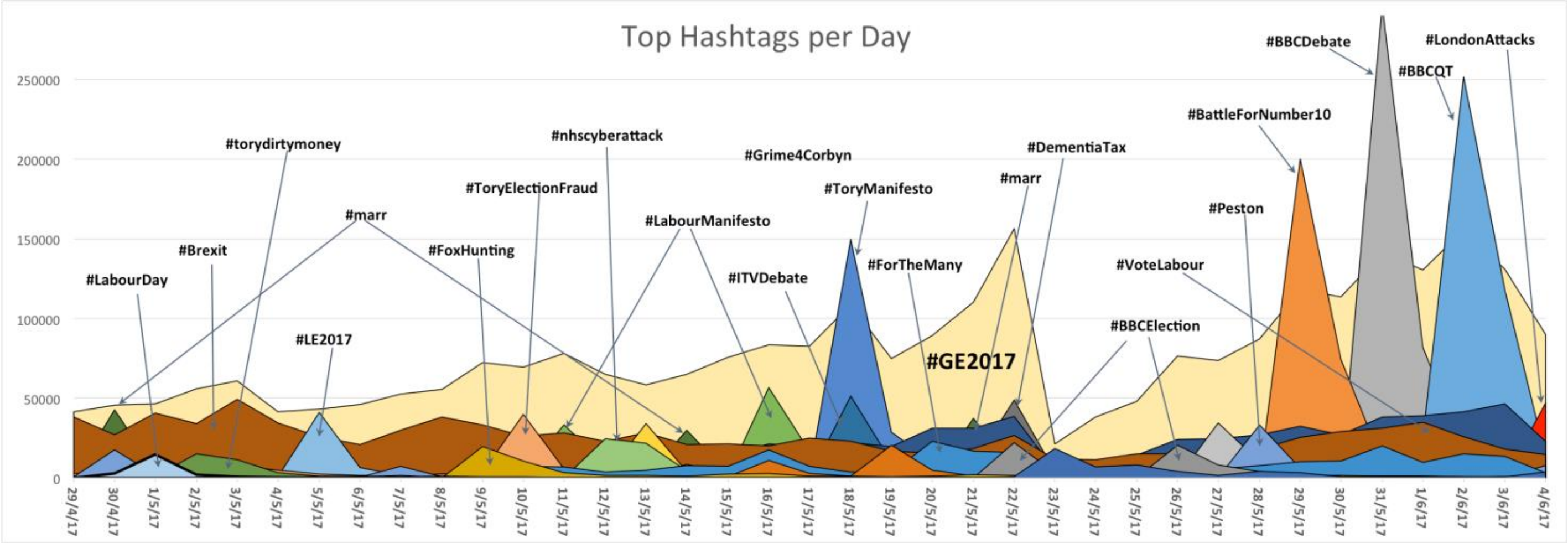

Figure 4. Hashtags that peaked on a given day on Twitter during the study period

## 3.2. Daily peaks in popular hashtag use

We also looked at frequency peaks of popular hashtags broken down by day over the collection period. This allows us to spot issues that may not have been tweeted about most overall but which also motivated people to comment in large numbers. The hashtags shown here appear in the top three on at least one of the days during the campaign. The event driven nature of Twitter is particularly obvious here. We can see that the largest peaks in the data come from debate-type events covered by traditional media, such as #BBCDebate, #BattleForNumber10 and #BBCQT. Similarly, other TV programmes also cause spikes in the data with #peston and #marr being particularly noticeable. We can also see that the Tory manifesto caused more of an impact that the Labour manifesto, possibly due to the leaking of the Labour manifesto, where we see two peaks of influence rather than one. We see the appearance of certain policy issues such as the so-called Dementia Tax and Fox hunting and in particular we see the sustained appearance of Brexit throughout the campaign. There are some issues which are notable by their absence as peak issues, for example the NHS and the economy. Also apparent are events that have taken place during the election campaign. Here we can see peaks of discussion of both the #NHSCyberAttack and the #LondonAttacks.

## 3.3. Most mentioned and retweeted accounts over the collection period

The story of Labour dominance in the Twittersphere continues when we examine the most retweeted accounts and the accounts that get the most mentions by others. Jeremy Corbyn tops both of these lists and, though Theresa May (654,417) is the second most mentioned account, she is mentioned only half as often as Corbyn (1,367,392). The difference between the two main party accounts @uklabour (323,027) and @conservatives (307,550) is much less. Both leaders are mentioned much more often than their respective parties, perhaps confirming the presidential tenor of the campaign. Striking once again is the disproportionate presence of the SNP (145,937) and this time also their leader Nicola Sturgeon (116,360) at positions six and seven in this UK wide debate. The SNP outperforms the Liberal Democrats (93,473) and their leader Tim Farron (69,009). Scottish Conservative and Unionist Party leader, Ruth Davidson (69,334), is also mentioned marginally more often than Tim Farron. UKIP (80,855) is the tenth most mentioned account, but again Paul Nuttal does not feature. The presence of the official media accounts, and also of polling agencies like Yougov, is again a prominent feature of the most mentioned accounts in relation to the GE2017 Twitter conversation. The most retweeted accounts were again heavily pro-Labour. We also saw, as we might expect, a strong presence of the professional media and of campaign bodies here. The generation of popular memes by @laboureoin also proved to be a very effective strategy for encouraging retweets carrying a socialist message.

Table 2: Top Retweeted and Top Mentioned Twitter Accounts

| Top retweeted accounts | | Top mentioned accounts | |
|---|---|---|---|
| **Account** | **Count** | **Account** | **Count** |
| @jeremycorbyn | 821,499 | @jeremycorbyn | 1,367,392 |
| @laboureoin | 426,912 | @theresa_may | 654,417 |
| @nhsmillion | 342,440 | @uklabour | 323,027 |
| @rachael_swindon | 290,587 | @conservatives | 307,550 |
| @owenjones84 | 224,605 | @bbcnews | 154,898 |
| @socialistvoice | 189,830 | @thesnp | 145,937 |
| @jeremycorbyn4pm | 186,269 | @nicolasturgeon | 116,360 |
| @davidjo52951945 | 178,768 | @skynews | 97,287 |
| @uklabour | 171,577 | @libdems | 93,473 |
| @davidschneider | 164,457 | @ukip | 80,885 |
| @jamesmelville | 162,626 | @thecanarysays | 80,176 |
| @chunkymark | 161,681 | @bbclaurak | 77,324 |
| @toryfibs | 146,873 | @ruthdavidsonmsp | 69,334 |
| @independent | 139,540 | @timfarron | 69,009 |
| @el4jc | 119,623 | @lbc | 65,571 |
| @britainelects | 115,655 | @yougov | 62,741 |
| @paulmasonnews | 108,302 | @borisjohnson | 61,292 |
| @imajsaclaimant | 108,130 | @guardian | 58,880 |
| @aaronbastani | 101,101 | @afneil | 52,216 |
| @peterstefanovi2 | 99,875 | @johnmcdonnellmp | 49,271 |

## 3.4. Daily peaks in most mentioned accounts

The top mentioned accounts were selected in the same way as the most frequent hashtags. All of these accounts appear in the top three most mentioned on at least one of the days during the campaign. Here we can see that @jeremycorbyn has been mentioned much more frequently than @theresamay throughout the campaign. We see three political parties mentioned; @conservatives, @uklabour and surprisingly again @theSNP. We can see how events shape peaks with the single peak in mentions for the Scottish Conservative and Unionist Party leader @ruthdavidsonmsp reflecting her announcement of a u-turn on prescription charges in Scotland. We can also see the influence of journalists here with @krishgm, @bbclaurak and @afneil.

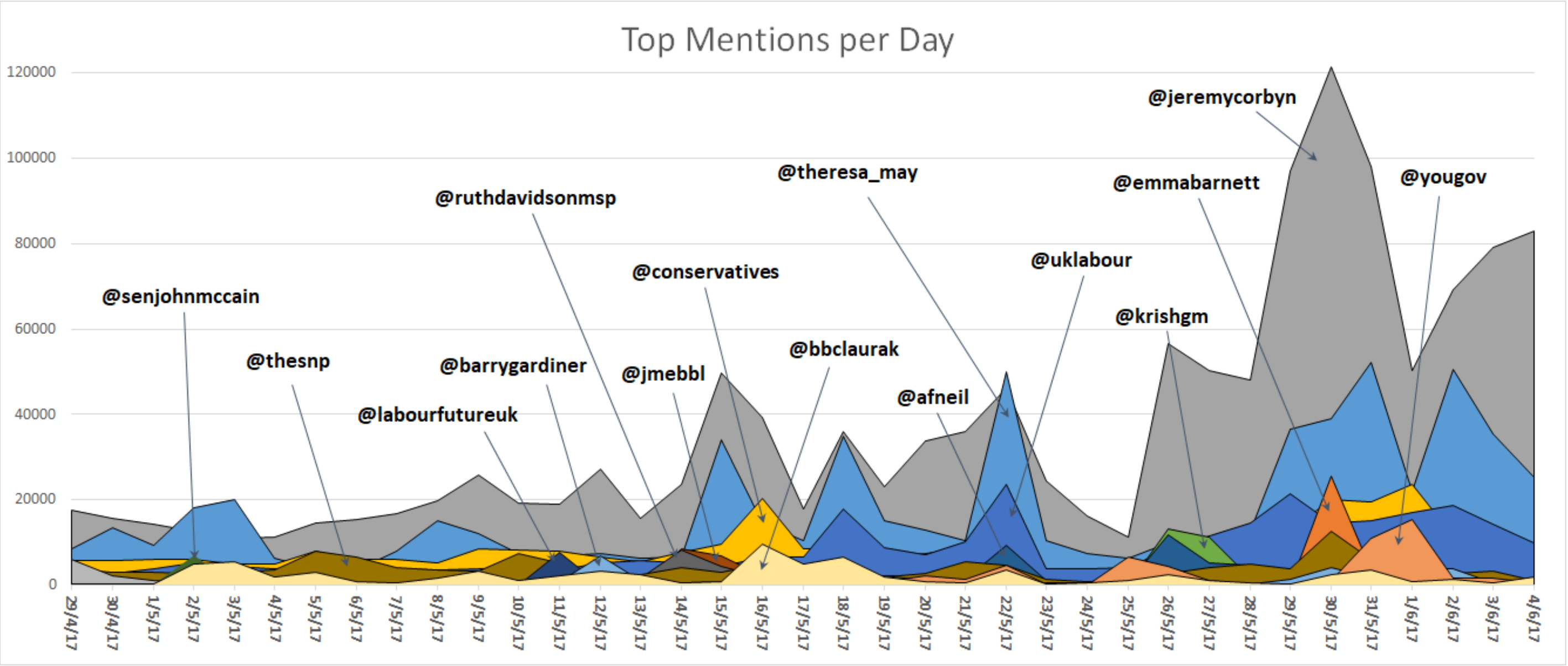

Figure 5. Mentions that peaked on a given day on Twitter during the study period

## 3.5. Most mentioned topics by and linked to politicians

We checked the most frequent terms used by key politicians themselves and we compare these with the most frequent terms used by others when mentioning these politicians. We can consider the terms used by the politicians as their attempts to influence debate and to set the agenda. This cloud reflects what the politicians want to talk about. The terms in the tweets from others mentioning the politicians can be considered to be the topics that Twitter users are trying to direct towards the politicians, these clouds reflect the agenda that Twitter users are associating with that politician.

In the visualisations you can see the terms side by side for each politician. There are several things that can be observed:

- Jeremy Corbyn is the only politician that directly challenges another politician - we can see this through the use of @theresamay. The other politicians do mention other leaders but they do not use specific @ mentions to interact with them;
- Theresa May and Tim Farron do not mention their parties often;
- Ruth Davidson who is often associated with distancing herself from the Scottish Tory brand actually tweets about her party quite often;
- Whereas Theresa May often tweets about Brexit this is not echoed in the tweets of those that mention her;
- Brexit does occur in the tweets mentioning Tim Farron, Nicola Sturgeon and Ruth Davidson;
- None of the Scottish leaders tweet heavily about independence but those tweeting about Kezia Dugdale and Ruth Davidson associate both of them heavily with this debate;
- Nicola Sturgeon, heavily associated in the traditional media with Scottish independence, is not associated with this by those mentioning her or seeking to interact with her on Twitter;
- The debates feature heavily in the tweets mentioning politicians except for Kezia Dugdale;
- Jeremy Corbyn and Nicola Sturgeon make most sophisticated use of Twitter devices such as hashtags and @ mentions.

Table 3. Tagclouds of top terms appearing with politicians own tweets v tweets mentioning them

<table>
<tr><th>Account own tweets</th><th>Tweets mentioning the account</th></tr>
<tr><td>
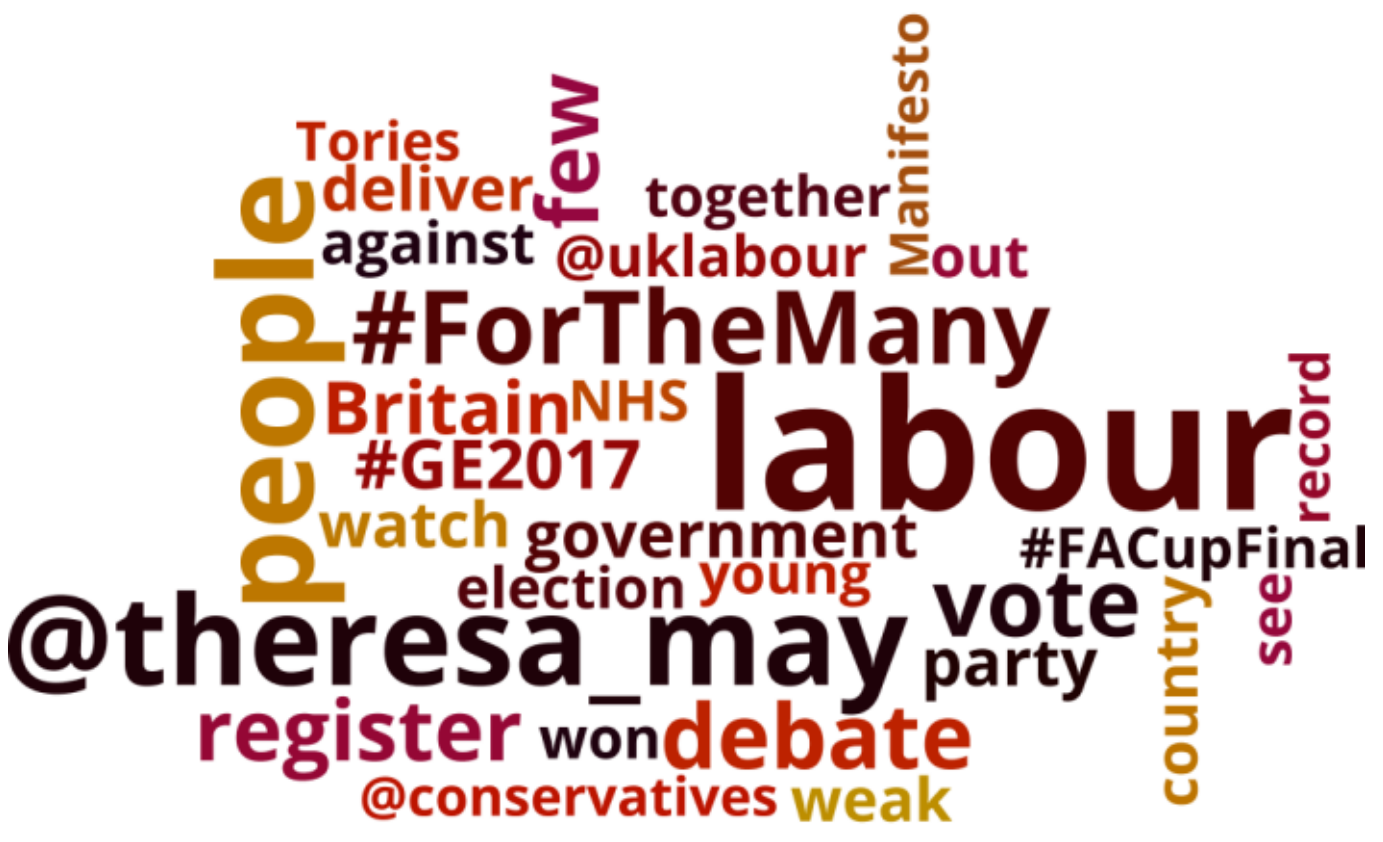

Jeremy Corbyn
</td><td>
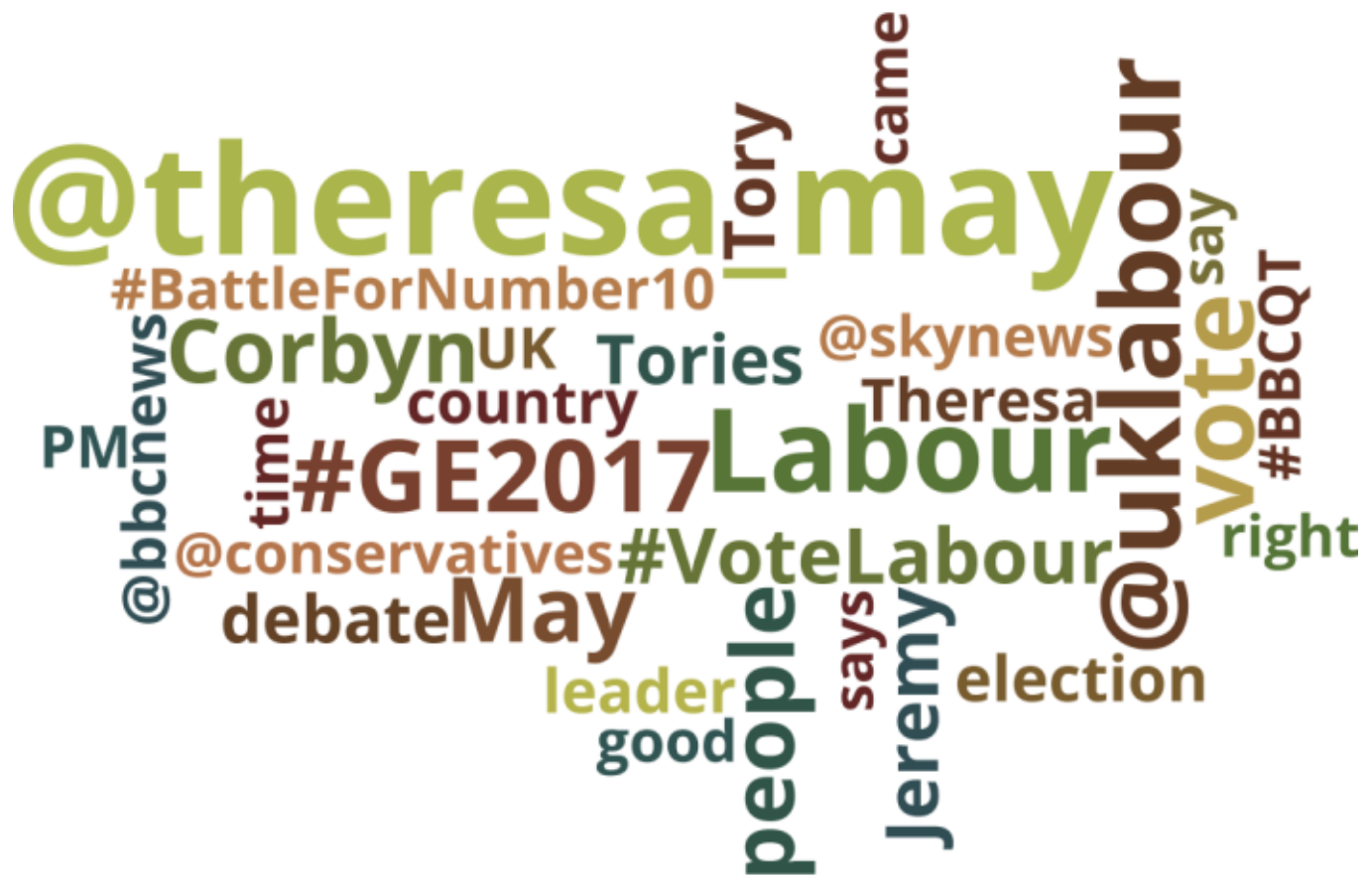

Jeremy Corbyn
</td></tr>
<tr><td>
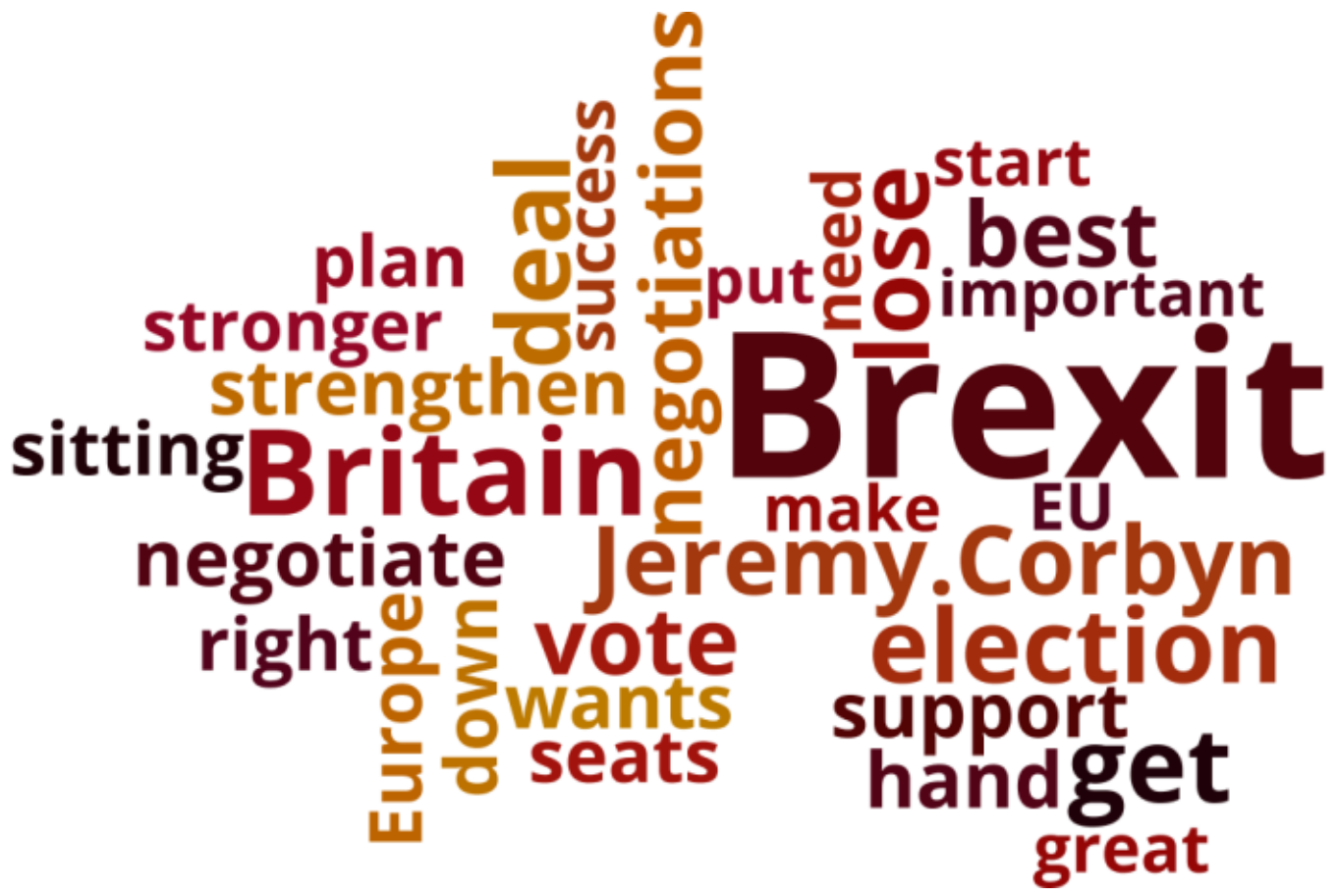

Theresa May
</td><td>
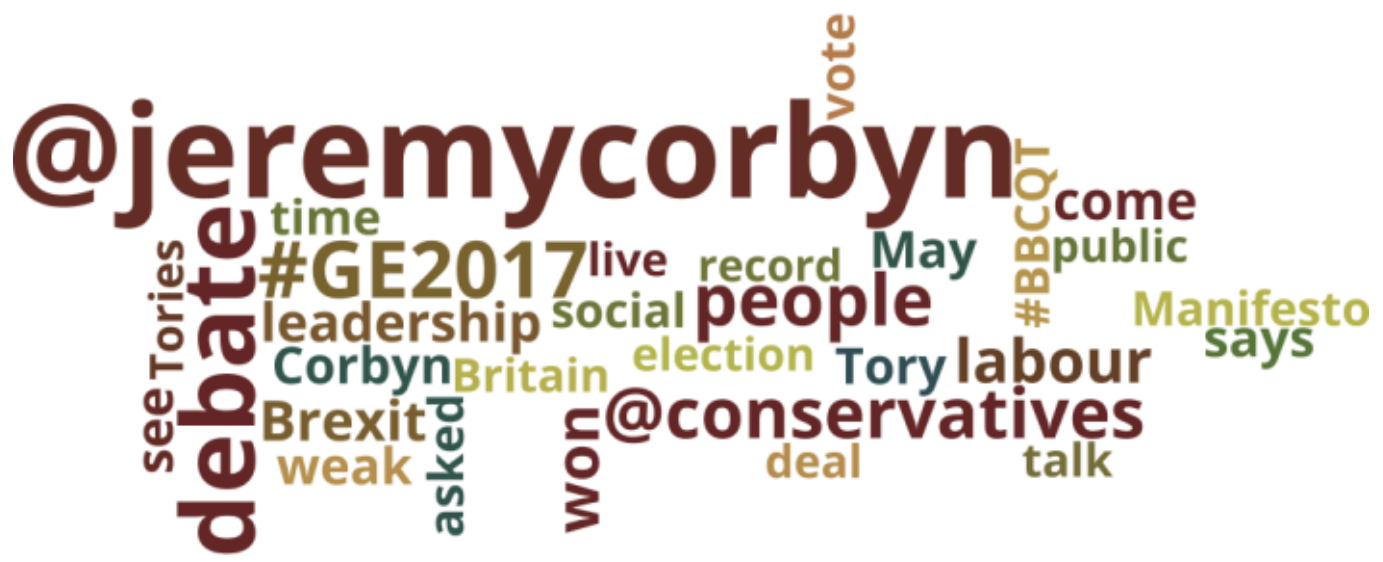

Theresa May
</td></tr>
<tr><td>
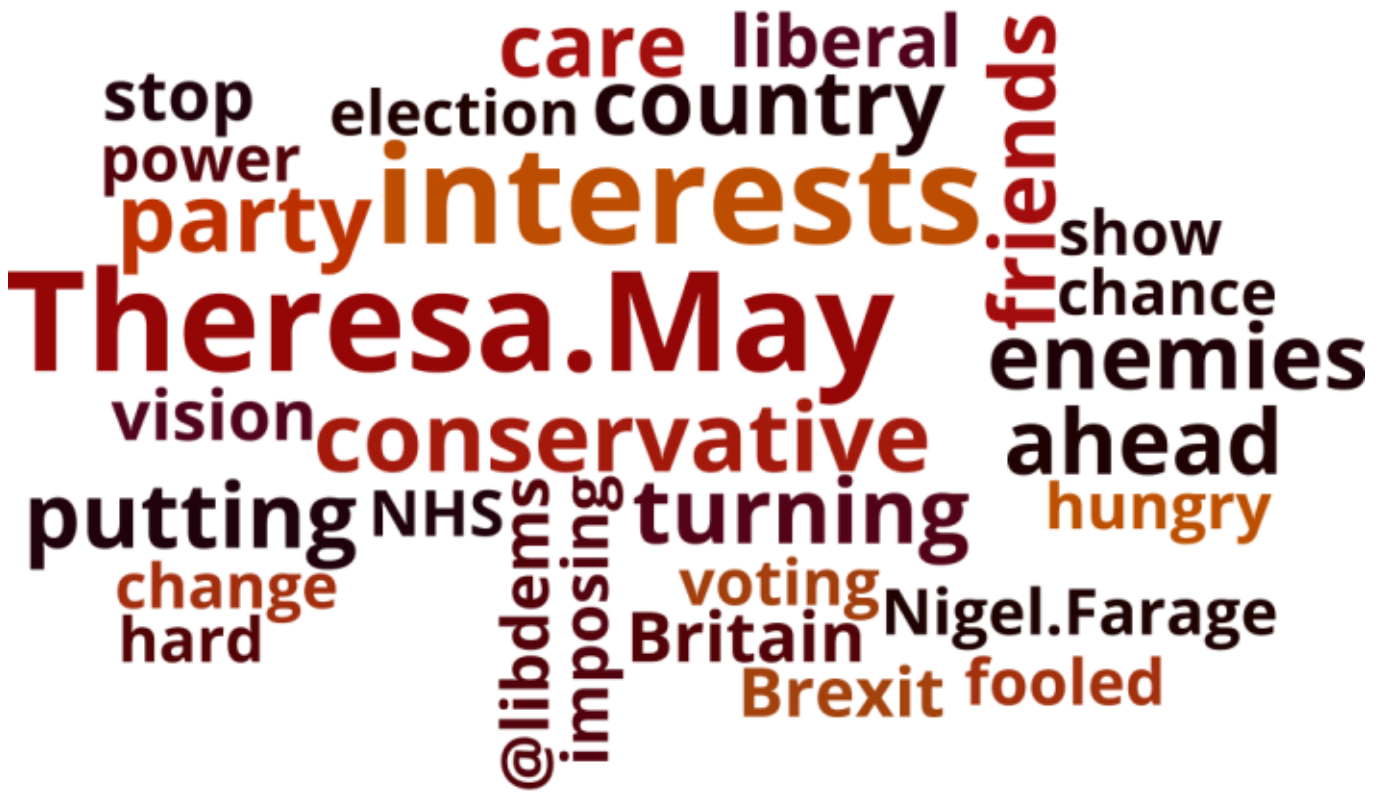

Tim Farron
</td><td>
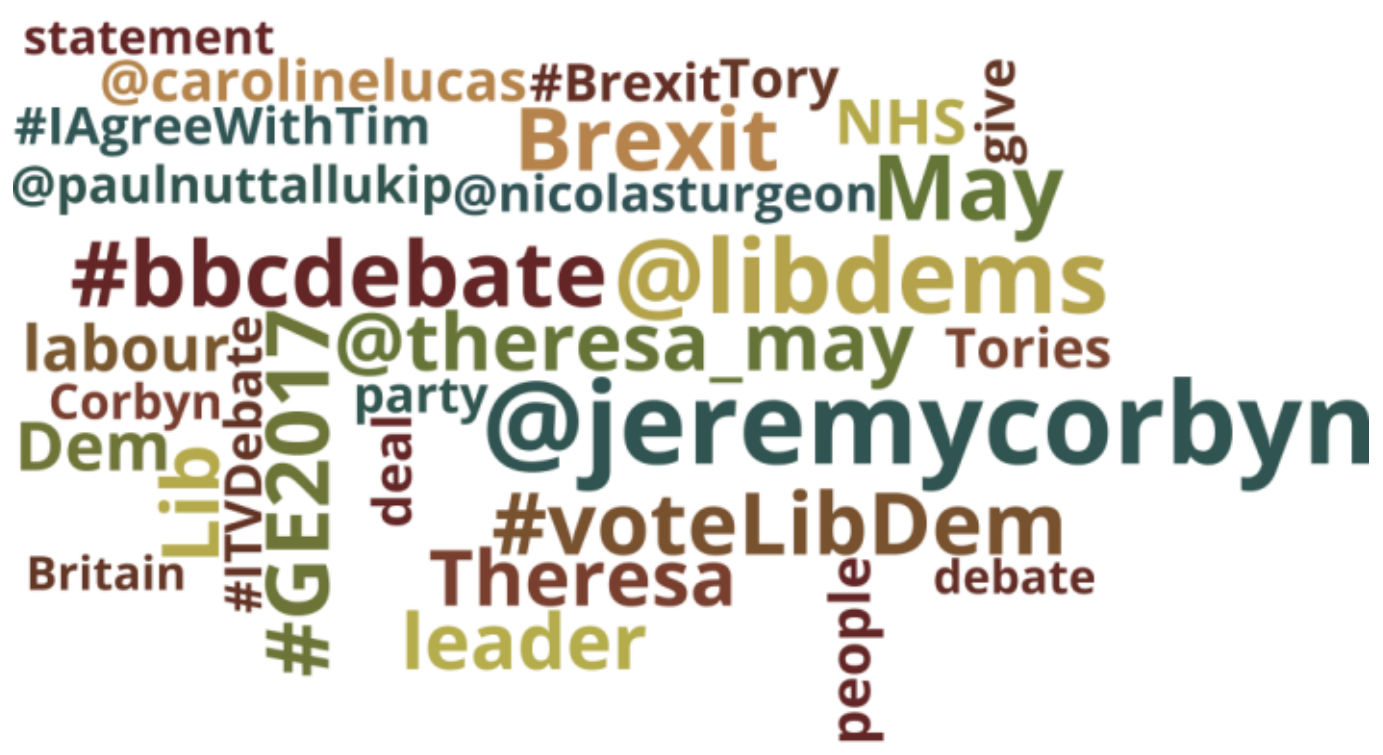

Tim Farron
</td></tr>
</table>

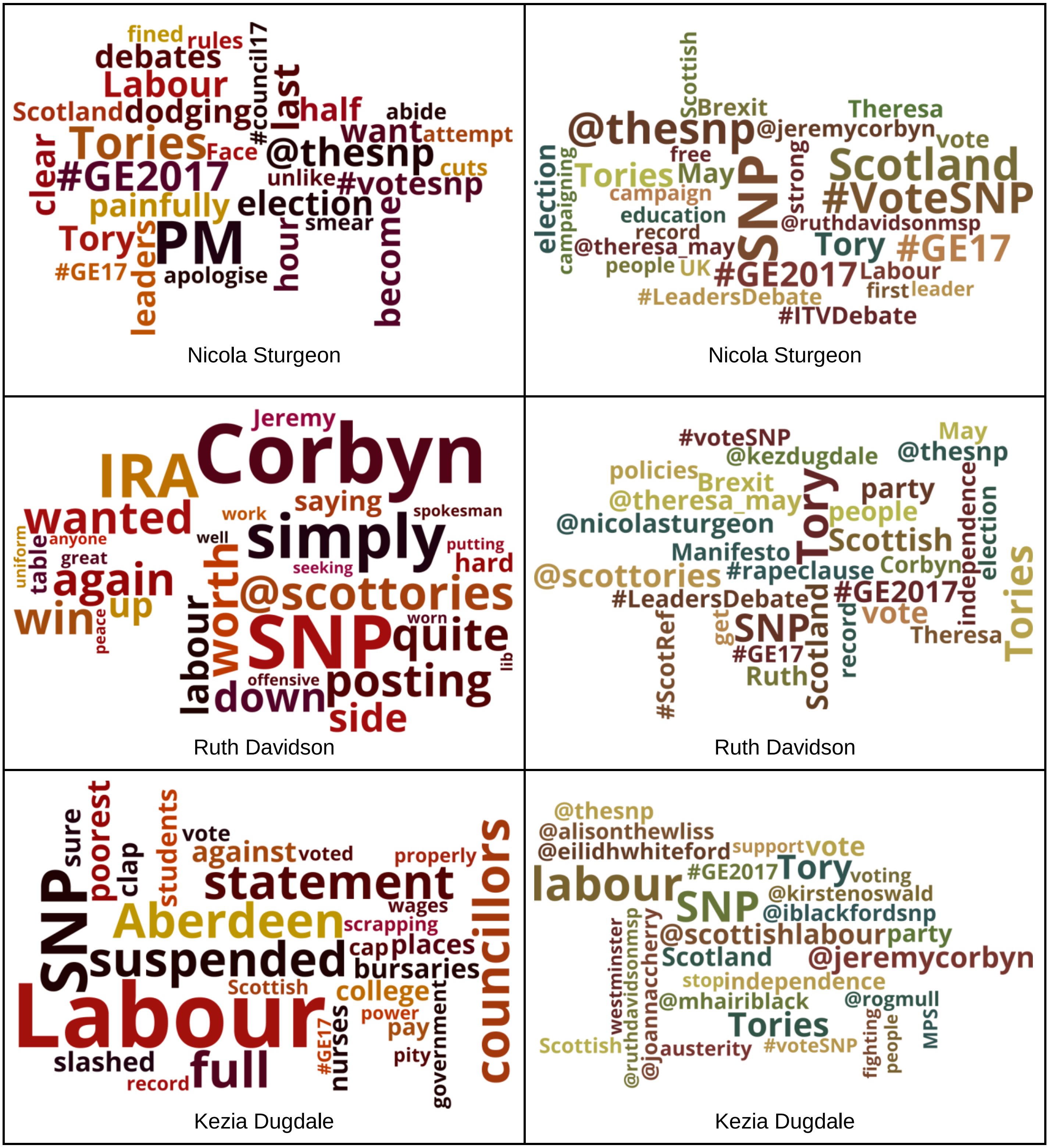

Figure 6: Most frequent terms employed by key politicians v the most frequent terms used by others when mentioning these politicians.

## 4. Discussion and Conclusions

In this study, we applied a quantitative analysis to the Twitter posts on the British general elections 2017 during the month period preceding the election. Our analysis included a set of around 35 million tweets on the topic. Here we present a preliminary exploratory analysis of this data set. Twitter analysis has strengths and weaknesses. Twitter users are not representative of the wider public – they are self-selected users not those chosen on the basis of careful sampling by opinion pollsters. Twitter users tend to be highly motivated (with an axe to grind), younger than average (though not exclusively young) [1] and are likely more often men [2] when engaged in political debate. So any insights are partial. That said, Twitter can be a reflection of spontaneous, motivated behaviour. Analysing Twitter narratives helps us to see where those highly motivated individuals position themselves in relation to the debate, what appears to provoke peaks in motivated activity and also what the overall trends are in these vocal and active publics. It also helps us to explore who sets agendas and shapes conversations in the Twittersphere and how effective or ephemeral these narratives are.

Events played a key role is shaping the Twitter conversation. These events can take many forms, election-specific events such as the debates, media events such as television and radio programs and physical events such as the NHS cyber attack and terrorist attacks. Twitter does not exist in a vacuum and the conversation that occurs there is often prompted by external events.

Researchers at the Centre for Research in Communication and Culture (CRCC) at Loughborough University have studied media coverage of the election campaigns over two weeks, 18th - 31st May. They analysed discussion in television news and print media. They observed that in the first week, starting on the 18th May, the Conservatives and Theresa May receive more coverage than Labour and Jeremy Corbyn. In contrast, we find that the social media discussion during this time was focused on Labour and Corbyn. The second week saw a rise in the traditional media coverage of the Labour party which was slightly more than the Conservatives. This rise was for the most part driven by increased traditional media coverage of Jeremy Corbyn bringing this more in line with the social media data. The Loughborough data also shows a high prevalence of Scottish MSP's with Nicola Sturgeon and Ruth Davidson both appearing in the top 10. This is mirrored in our social media data set.

In the last week of May the top five most prominent issues seen in television and print media were the electoral process, Brexit, health care, taxation and the economy/business. We can see this echoed to some extent in the social media data, with the manifestos, Brexit, and the NHS. We do not see the appearance of taxation, the economy or business issues.

Our data set shows an overwhelming dominance of pro-Labour tweeting. With over 1 million tweets using Labour hashtags, Labour party coverage out-performs pro-Conservative coverage by almost three times. There is a disproportionate presence of the SNP in this social media set and given that only Scottish voters can elect this party this was particularly striking. These three observations were also noted in a week long study (1-7th May) conducted by researchers at the Oxford University's Internet Institute [4].

It is important not to exaggerate the novelty of new social media. New media to some extent appears to be simply an extension of old media and we see in our data set how broadcast media often generates additional coverage by effectively reporting on itself. Key bursts in Twitter throughout the GE2017 campaign came from people using hashtags associated with TV debates and programmes like Question Time. These hashtags are created by the programme makers and die off almost immediately. This is not surprising as they are instantaneous or throw-away hashtags associated with a specific programme. It would be difficult to claim that these had any significant role in setting a political agenda or in shaping debates, rather they act as a means of tracking who was watching and engaging with the particular media generated events.

Twitter is of course not representative of the voting public as a whole, and therefore not necessarily a clear reflection of “the many, not the few”. However, whilst Twitter cannot be used to predict elections and the overwhelming support we see for Labour and Jeremy Corbyn may not be fully reflected in the ballot boxes, it is a useful tool in giving us the mood of those who are motivated enough to comment in social media. Tweeters are typically highly motivated and perhaps those who initially see themselves as the underdogs in the debate, excluded from mainstream coverage. This has been apparent in a number of recent campaigns. The YES campaign, though ultimately unsuccessful, dominated social media in the Scottish independence referendum. Leave groups did the same in the 2016 Brexit campaign and Trump’s dominance in social media transformed US election coverage, with both these campaigns ultimately triumphing at the polls. As the Loughborough study shows, Corbyn’s campaign did not initially enjoy the access to the traditional media that May was afforded [3]. This may explain the surge in social media activity which subsequently developed a life of its own and has ultimately had to be acknowledged by the mainstream media. This also fits with the high presence of the SNP in our data set, with the Scottish debate marginalised at the UK level. If the current polling is to be believed Jeremy Corbyn is unlikely to do as badly as was anticipated when the election was first called. Traditional media sources were slow to pick up on this change in public opinion whereas this trend could be seen early on in social media and throughout the month of May.